\documentclass[fleqn,10pt]{format} 

\usepackage[english]{babel} 

\usepackage{lipsum} 

\usepackage{float}

\definecolor{color1}{RGB}{0,0,90} 
\definecolor{color2}{RGB}{0,20,20} 

\usepackage{hyperref} 
\hypersetup{hidelinks,colorlinks,breaklinks=true,urlcolor=color2,citecolor=color1,linkcolor=color1,bookmarksopen=false,pdftitle={Title},pdfauthor={Author}}

\JournalInfo{} 
\Archive{} 

\PaperTitle{A Practical Guide to Spectrogram Analysis for Audio Signal Processing} 

\Authors{Zulfidin Khodzhaev\textsuperscript{1}} 
\affiliation{\textsuperscript{1}\textit{Department of Electrical and Computer Engineering, University of Texas at Austin, Austin, TX 78712, USA}}

\newcommand{\keywordname}{Keywords}

\Abstract{The paper summarizes spectrogram and gives practical application of spectrogram in signal processing. For analysis, finger-snapping is recorded with a sampling rate of 441000 Hz and 96000 Hz. The effects of the number of segments on the Power Spectral Density (PSD) and spectrogram are analyzed and visualized.}

\begin{document}
\flushbottom 
\maketitle

\tableofcontents 
\thispagestyle{empty} 


\section*{Introduction} 

\addcontentsline{toc}{section}{Introduction} 
Dynamic signals such as output signals coming out of a communication device vary in their amplitude with time. Dynamic signals have two main problems namely, analyzing them, especially spectral analysis, and processing the signal i.e. designing filters to transform signals in the way we want. To analyze time-varying signals coming from devices, a waveform must be established to visualize and transform the data into a useful form. The waveform can be generated by placing the amplitude of a signal on the vertical axis and the time on the horizontal axis. The spectral analysis of a signal i.e. looking through the frequency content of the signal. 

To get the spectral analysis of a signal, the signal first decomposed into a sinusoidal signal such as:

\begin{equation}
   x(t) = A_0 + \sum_{k=1}^{N}(A_{k}\cos[2\pi f_kt + \phi_k)],
\end{equation}
where $A_0$ is constant, $A_k$ is amplitude of the signal, $f_k$ is frequency of the signal, $\phi_k$ is phase of the signal. 

For example, if the signal is:

\begin{equation}
x(t)=Acos(2\pi f_kt)
\end{equation}

Then, using the Euler equation, the cosine function can be rewritten as:

\begin{equation}
cos(2\pi f_kt)=\frac{e^{i2\pi f_kt} + e^{-i2\pi f_kt}}{2}
\end{equation}

and the signal is turned into a complex exponential signal and is decomposed as:

\begin{figure}[h]
\centering
\includegraphics[scale=0.6]{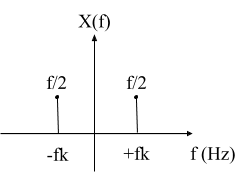}
\caption{Fourier transform of the cosine function. }
\label{Figure1}
\end{figure}

To best approximate a signal, the signal is approximated by a finite sum of sinusoids:

\begin{equation}
   x(t) = \sum_{k=-N}^{N}(a_{k}e^{i2\pi f_kt)},
\end{equation}
where $a_k$ is the complex number and is amplitude and the phase of the sinusoids, $f_k$ is the phase of the signal. 

When we have any signal coming from the device, it can be decomposed into a sinusoidal signal and for different frequencies, the $a_k$'s of sinusoids will be found. Consequently, the spectral analysis's idea is to find $a_k$'s of the signals.

In an example of periodic square signals, the signal is decomposed using sinusoids, and the location, frequency, and $a_k$'s of sinusoids are defined. The process of reconstructing square signals using sinusoids gives the exact form as the original signal itself when the relevant values of sinusoids is found and the example of this process can be seen in Figure 2.

\begin{figure}[h]
\centering
\includegraphics[scale=0.6]{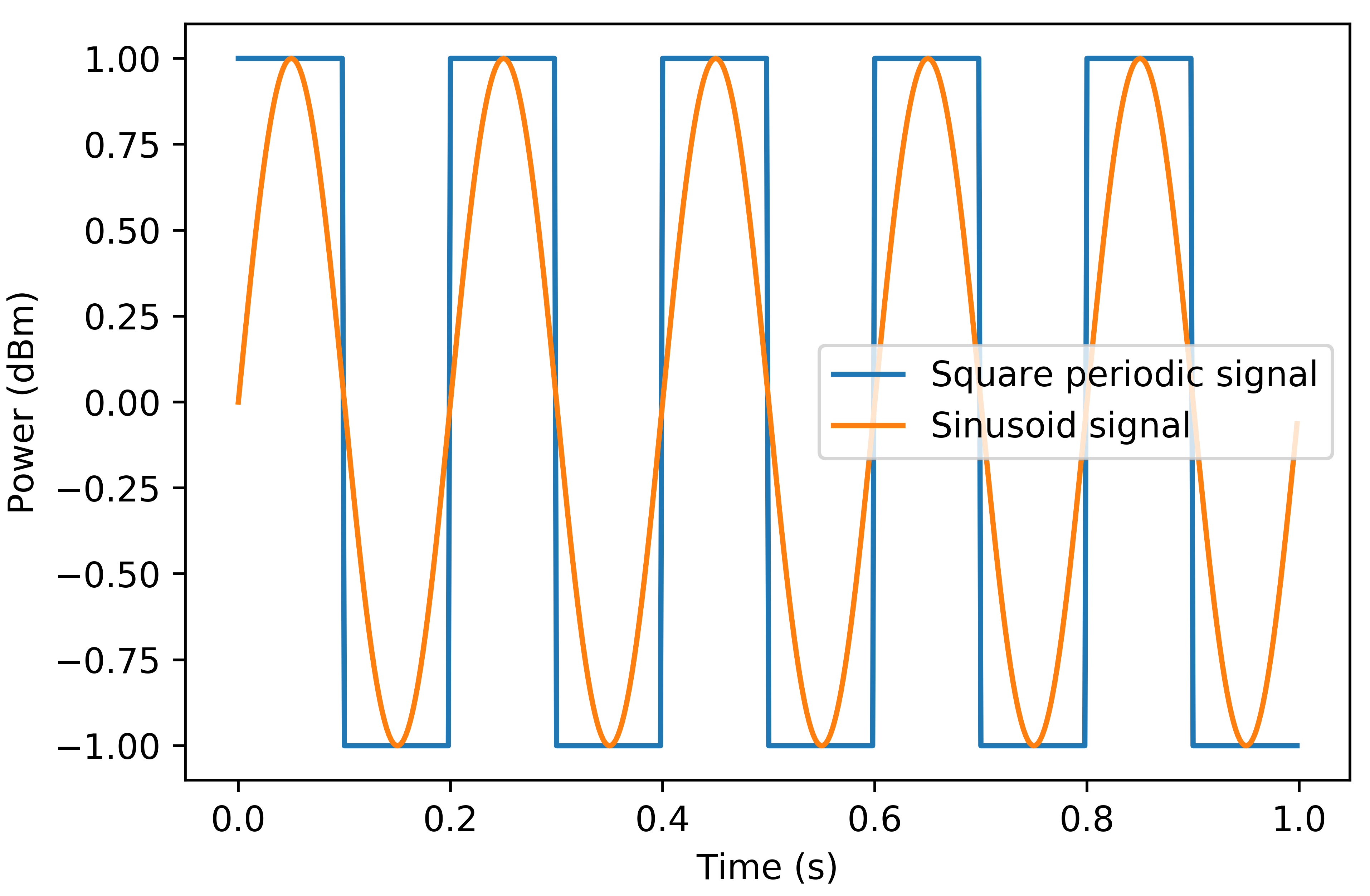}
\caption{Square periodic and sinusoid signals. }
\label{Figure2}
\end{figure}

By adding more sinusoids, the approximation becomes closer to the square periodic wave which can be seen in Figure 3 and the best approximation can be seen in Figure 4.

\begin{figure}[h]
\centering
\includegraphics[scale=0.6]{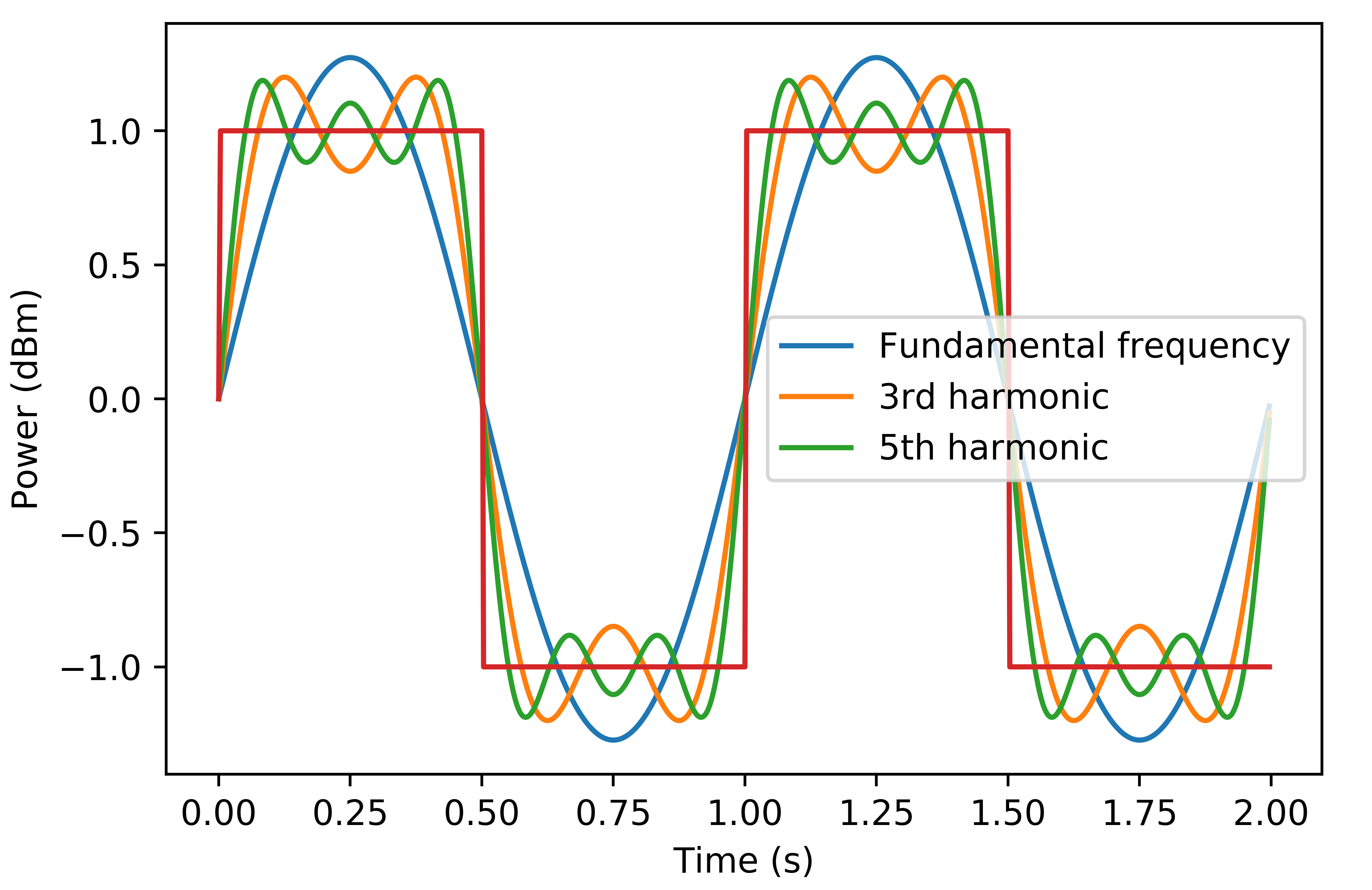}
\caption{Different harmonic sinusoids.}
\label{Figure3}
\end{figure}

\begin{figure}[h]
\centering
\includegraphics[scale=0.6]{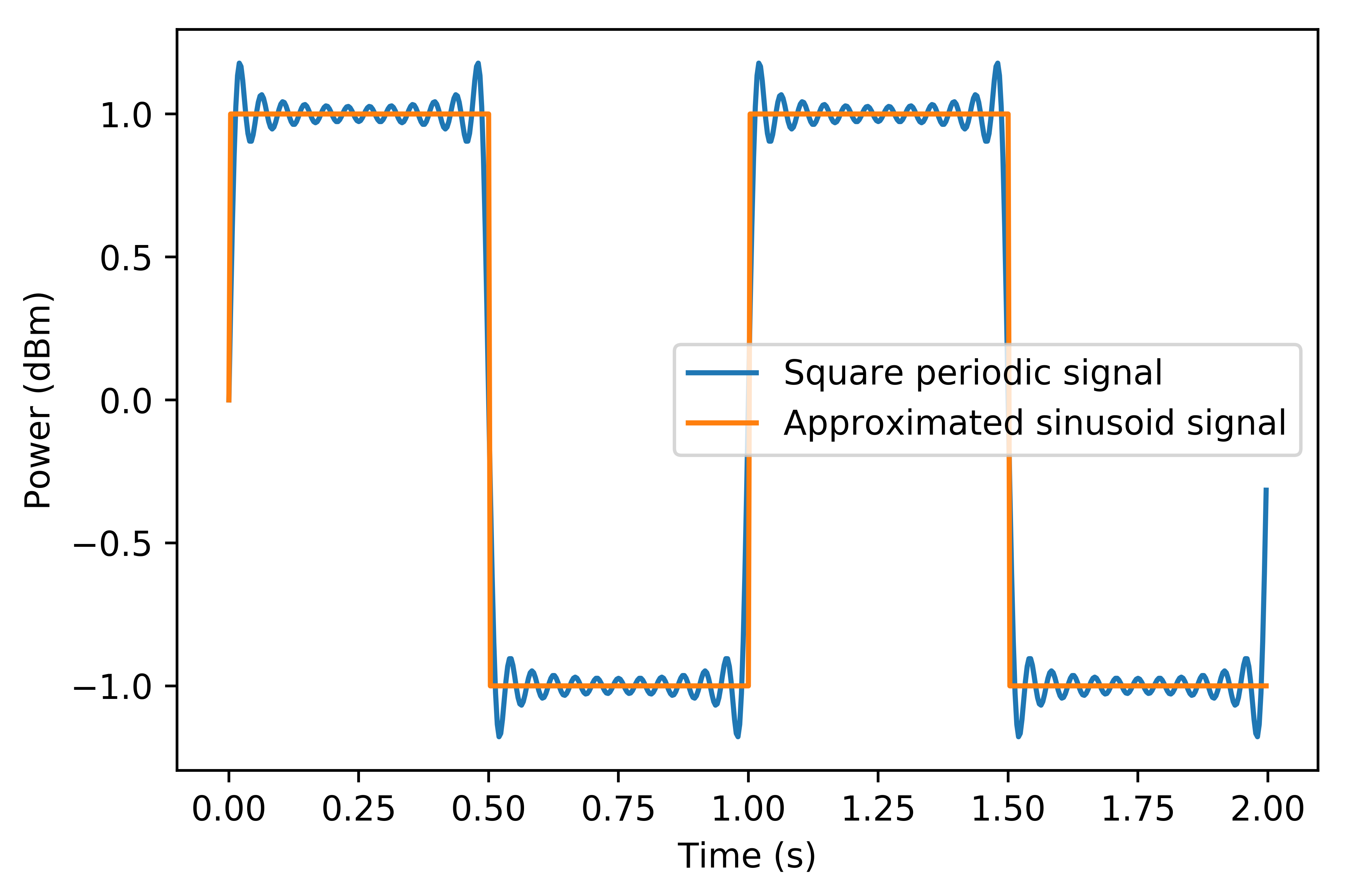}
\caption{Sinusoid approximation for square periodic wave.}
\label{Figure4}
\end{figure}

Spectral analysis of the periodic signals of the square wave is executed due to its frequency being the same. When the frequency content of the signal changes, two-dimensional spectral analysis of two dimensional plots will not be sufficient due to changes in time. There needs to be time involved in the spectral analysis process i.e. for each point in time, the spectral analysis needs to be performed which is called Time Frequency spectral analysis (TFSA).

The process of spectral analysis in time is spectrograms and this tool will help to understand the frequency content of the signal. As an example, the function $x(t)=cos(150\pi t + 9000\pi t^2)$ increases in frequency as time progresses, and the increase in frequency can be seen in Figure 5.

\begin{figure}[h]
\centering
\includegraphics[scale=0.6]{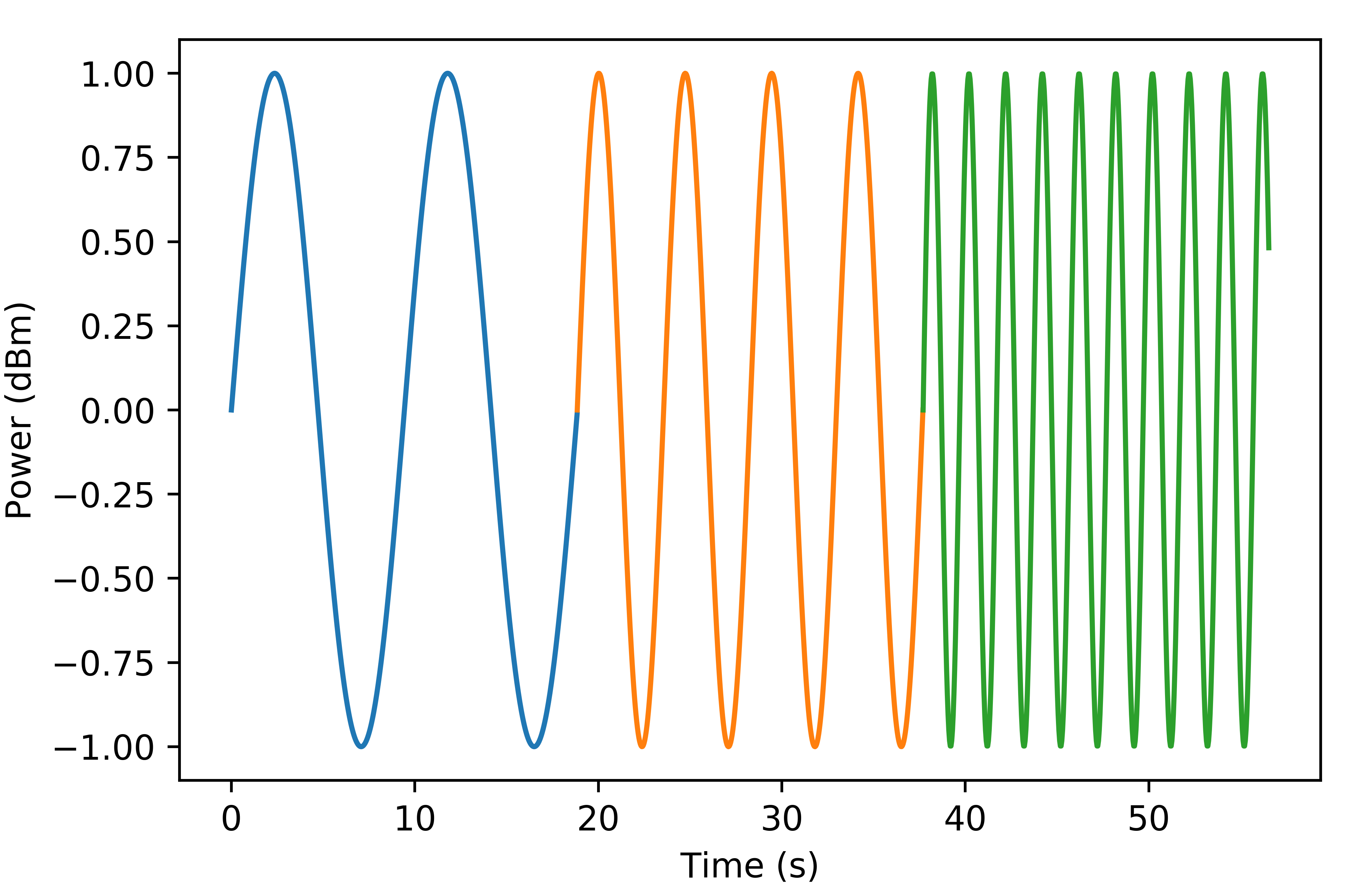}
\caption{Frequency increase in time.}
\label{Figure5}
\end{figure}

The frequency increase in time's graph will give the spectrum graph when it is graphed versus time. The spectrum graph of the frequency increase when time changes, can be seen in Figure 6.

\begin{figure}[h]
\centering
\includegraphics[scale=0.6]{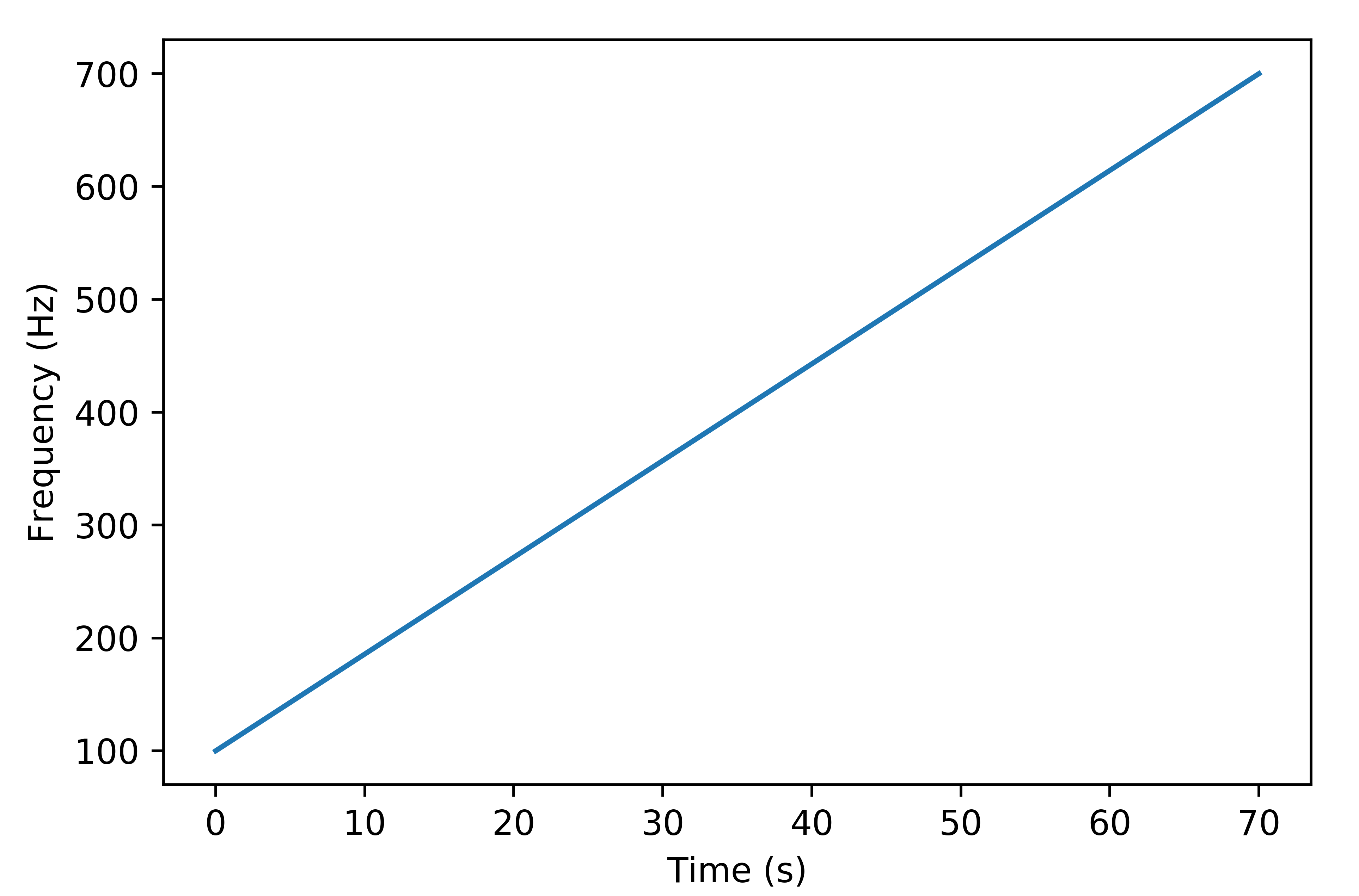}
\caption{Frequency vs Time.}
\label{Figure6}
\end{figure}

On the other hand, most of the real-world signals can not be expressed by mathematical expressions. For this purpose, for the data obtained from actual measurements or data obtained by using samples, Digital Fourier Transform (DFT) or Fast Fourier Transform (FFT) can be used for practical applications.

To see the color of the spectrogram which indicates the magnitude of the Power Spectral Density (PSD), the signal will be divided into sections. The sections can be arbitrary and in spectrogram, it can increase the resolution when the number of samples included is increased for Fourier Transform.

In this paper, PSD is obtained using Welch's average periodogram (WAP) method \cite{welch1967use}. WAP takes the squared magnitude of the FFT and averages the result. For more in-depth formulation of FFT \cite{4472244}\cite{bartlett1950periodogram} and periodogram \cite{oppenheim_schafer_2014}.

Obtained FFTs are also windowed, in this paper, using Hanning Windowing.
When the signal's frequency component is obtained using the FFT algorithm, the algorithm assumes that the signal is finite and that it is a periodically continuous signal with the connection between its time waveform and frequency domain. And when the signal has an integer number of periods with periodicity, the FFT calculates it correctly. If the signal is not an integer number of periods, then windowing will reduce the amplitude of the discontinuities around the borders of the samples. 

Time versus Frequency and the complex amplitude of those Fourier series will give the spectrogram. And magnitude of power spectral density will give the color of the spectrogram to understand the power of the content of the signal \cite{oppenheim1999discrete}.


\section{Spectrogram of finger snapping with 44100 Hz sampling rate}

\subsection{Frequency of the signal per time.}

To illustrate the process of finding a spectrogram of the signal with different frequencies, snaps were recorded using an audio recorder. The snaps were performed by a male of 24 years old in a library room. The sound was recorded for a total of 2 seconds, and the first fingers were snapped around 0.25 seconds and the second and third snap was snapped around 1 and 1.5 seconds respectively.

Initially, the finger-snapping was recorded with a sample rate of 96000 Hz - high-quality sound considering typical CD-quality audio has 44000 Hz samples per second. The record is plotted in Figure 7. 

\begin{figure}[h]
\centering
\includegraphics[scale=0.5]{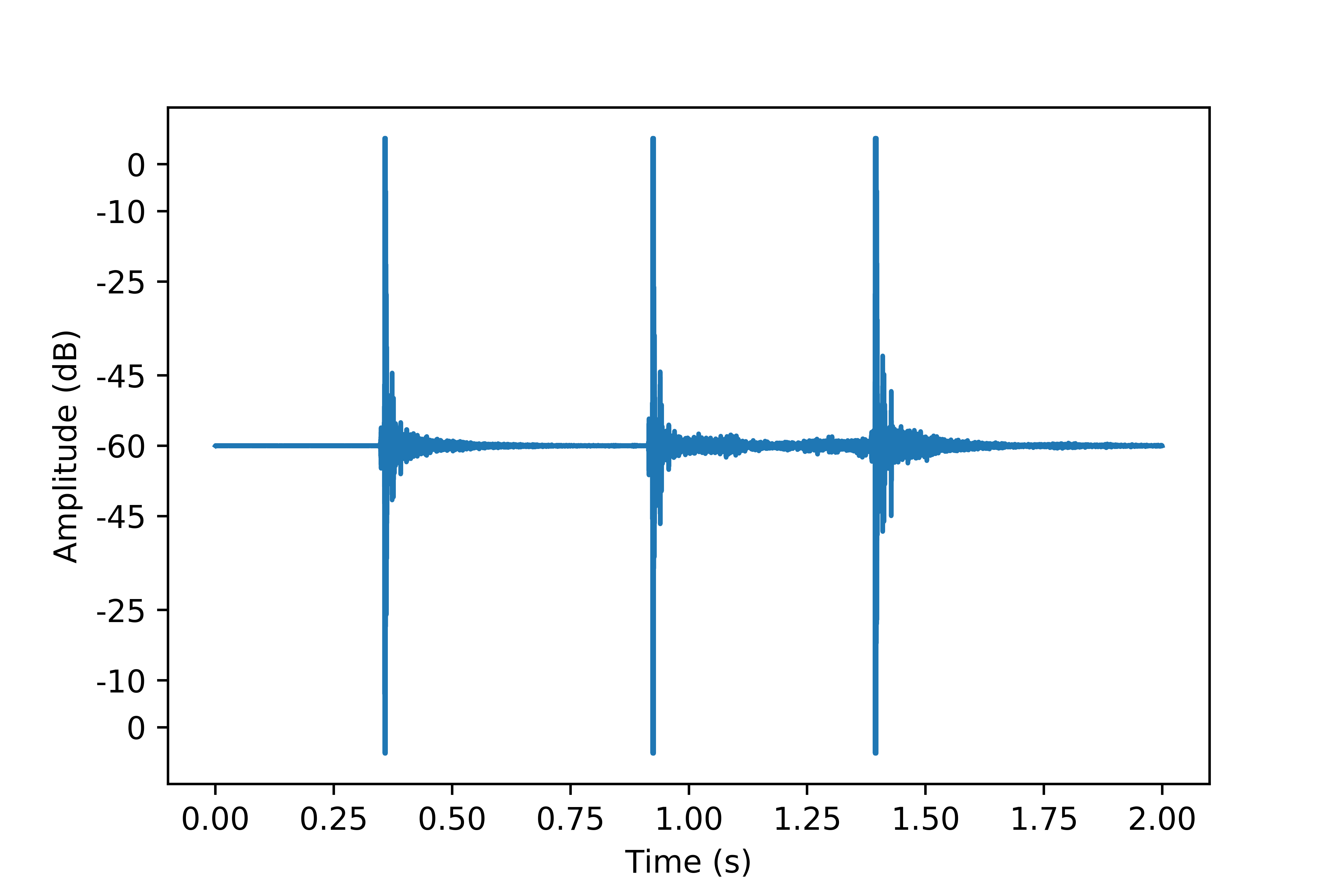}
\caption{Finger snapping signal graph.}
\label{Figure7}
\end{figure}

\subsection{Power Spectral Density}
 The power spectral density was plotted using  Welch's average periodogram method.  The signal is divided into 256 length parts and each segment is
windowed Hanning window. The overlap between segments was reduced to zero.

Then, the Power spectral density (PSD) versus frequency was graphed to give the information on the spectrogram and it is shown in Figure 8. From Figure 8, it can be seen that the PSD of the signal around 4500 Hz is the highest and it declines as frequency increases which is what is expected.

\begin{figure}[h]
\centering
\includegraphics[scale=0.5]{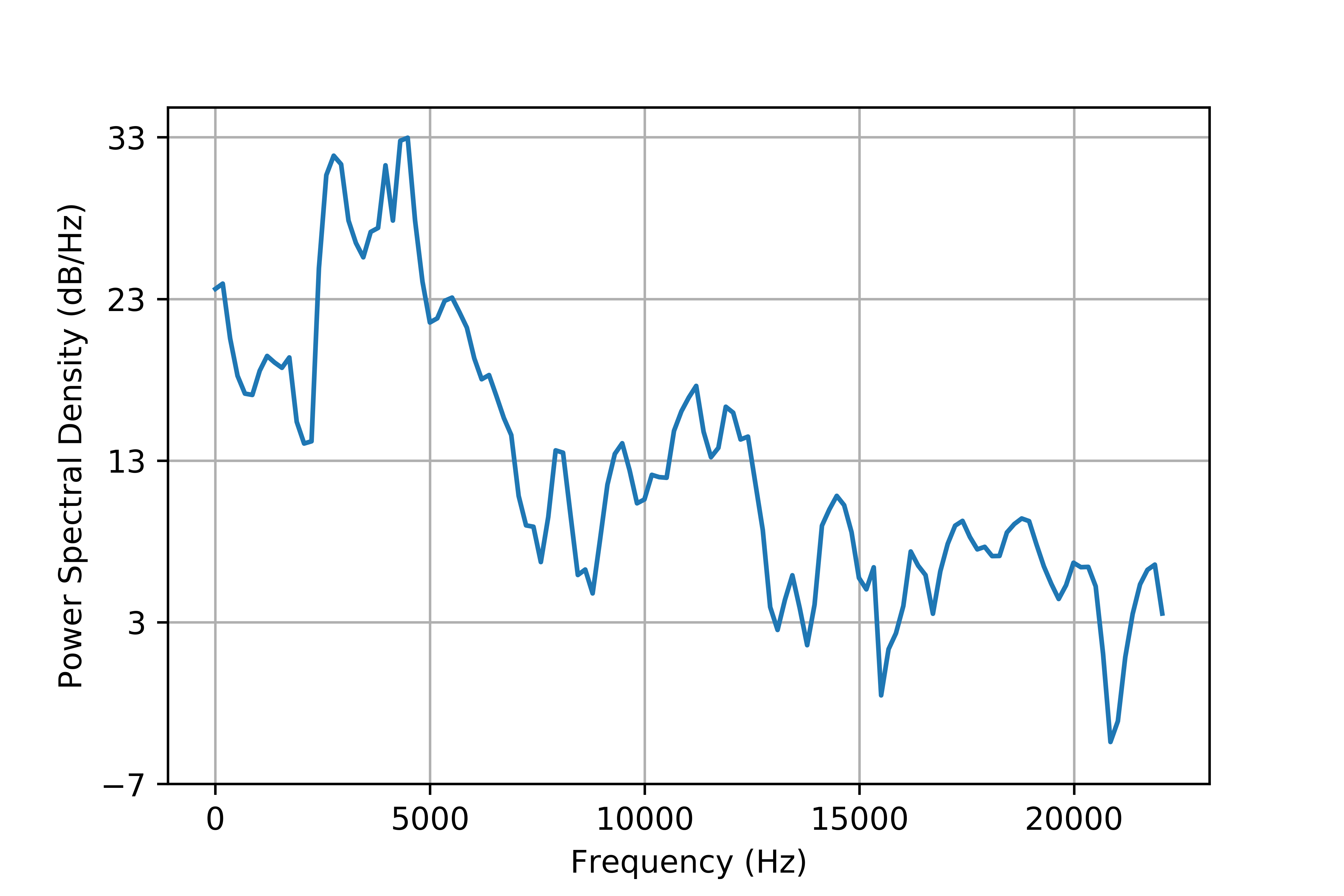}
\caption{Power spectral density versus frequency.}
\label{Figure8}
\end{figure}

\subsection{Spectrogram}

The signal was divided into 256 segments and each segment's FFT was calculated. Each segment was windowed by Hanning window and between each segment 128 overlaps were allowed.

The spectrogram of the signal is graphing time versus frequency and shows the magnitude of PSD of the signal by color which is shown in Figure 9.

\begin{figure}[h]
\centering
\includegraphics[scale=0.5]{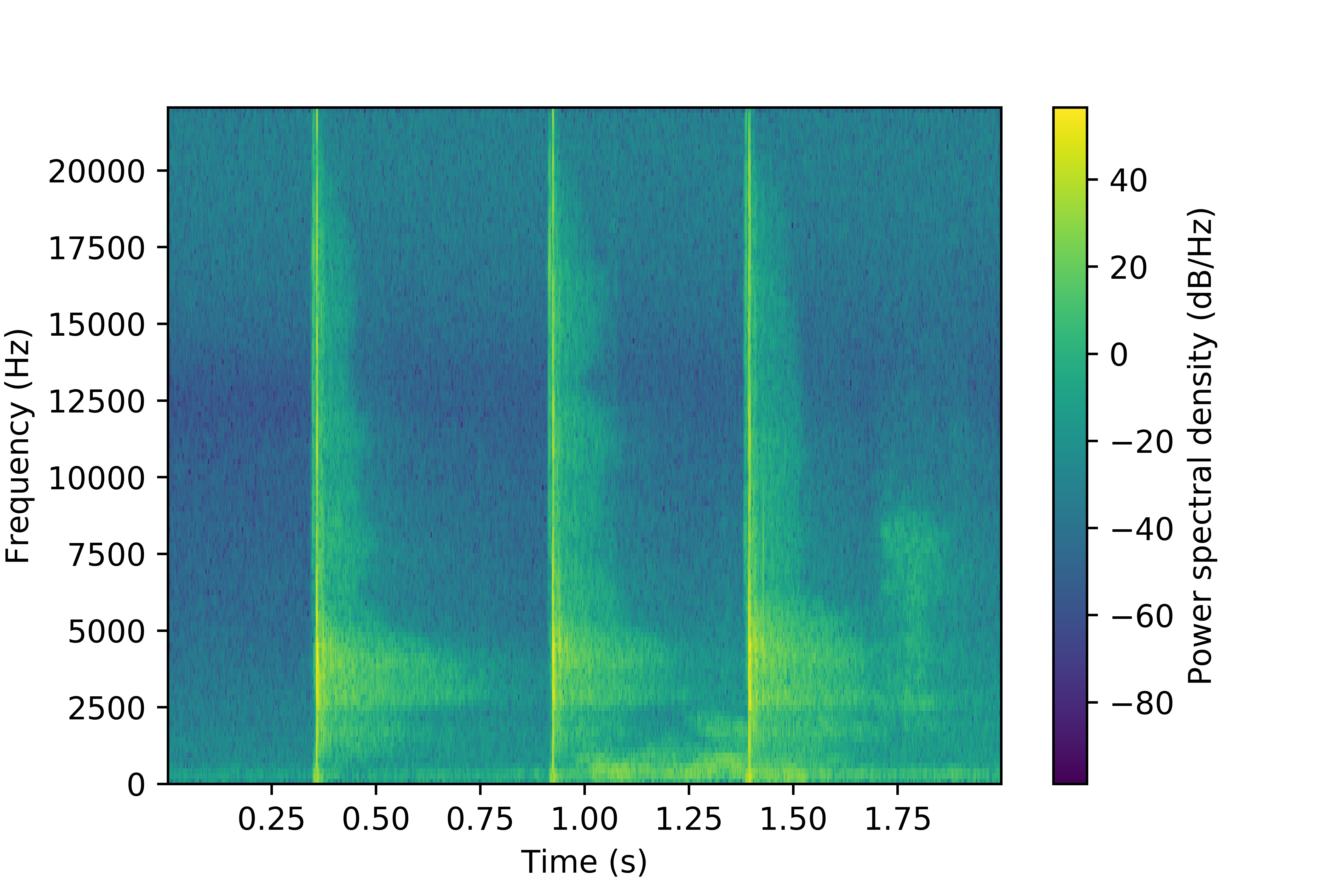}
\caption{Spectrogram of the finger snapping.}
\label{Figure9}
\end{figure}

From Figure 9, it can be seen that the PSD of the signal increases around 4500 Hz which is what was observed when Figure 8 was graphed. 

\section{Spectrogram of finger snapping with 96000 Hz sampling rate}

A male performed the finger-snapping in a room inside the library. The snaps were recorded for 2 seconds, with the first snap's time interval being longer than the second and third snaps. The sampling rate for the signal was increased to 96000 Hz. The signal is shown in Figure 10, the power spectral density in Figure 11, and the spectrogram in Figure 12.

\begin{figure}[h]
\centering
\includegraphics[scale=0.5]{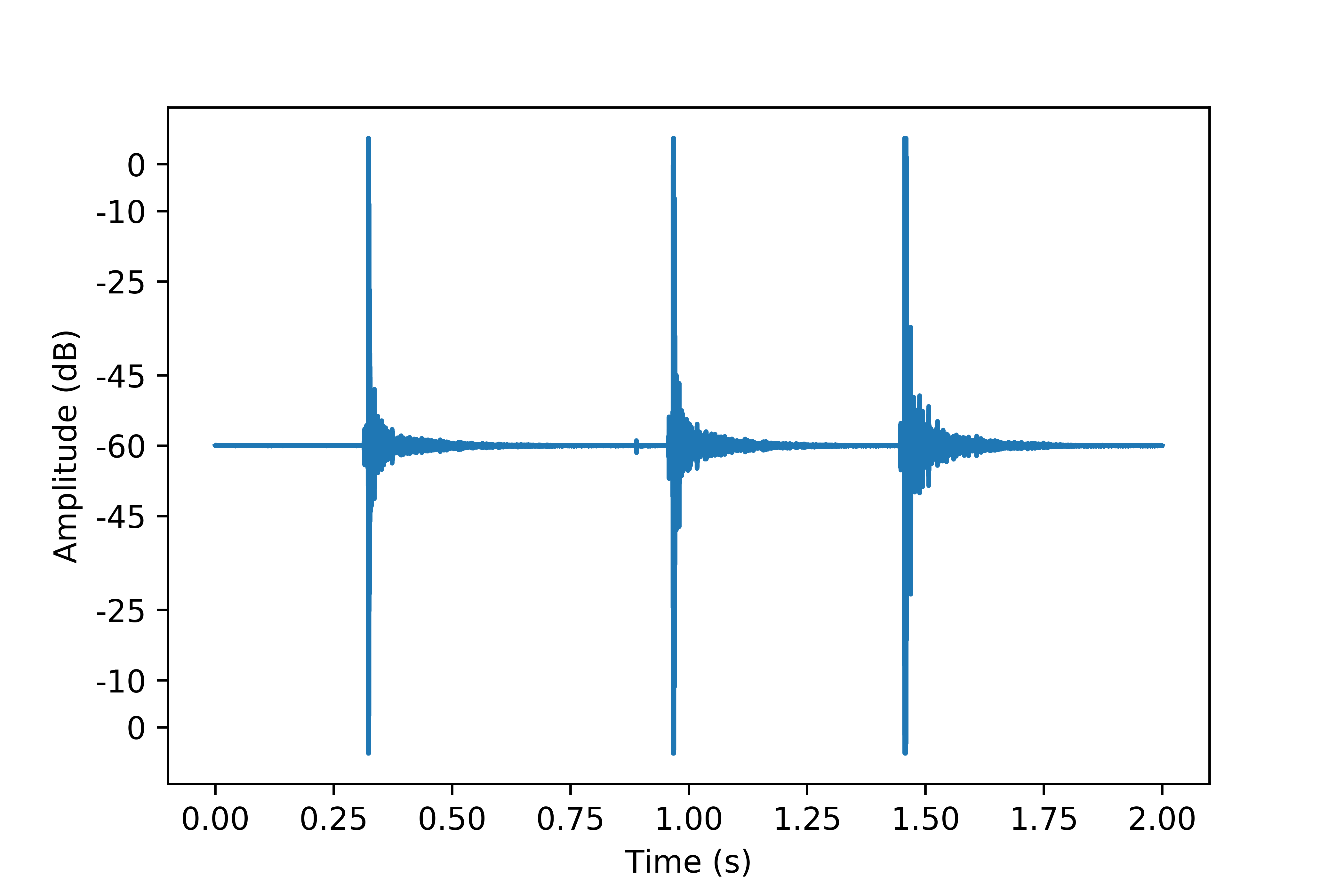}
\caption{Finger snapping signal graph with 96600 Hz sampling rate.}
\label{Figure10}
\end{figure}

\begin{figure}[h]
\centering
\includegraphics[scale=0.5]{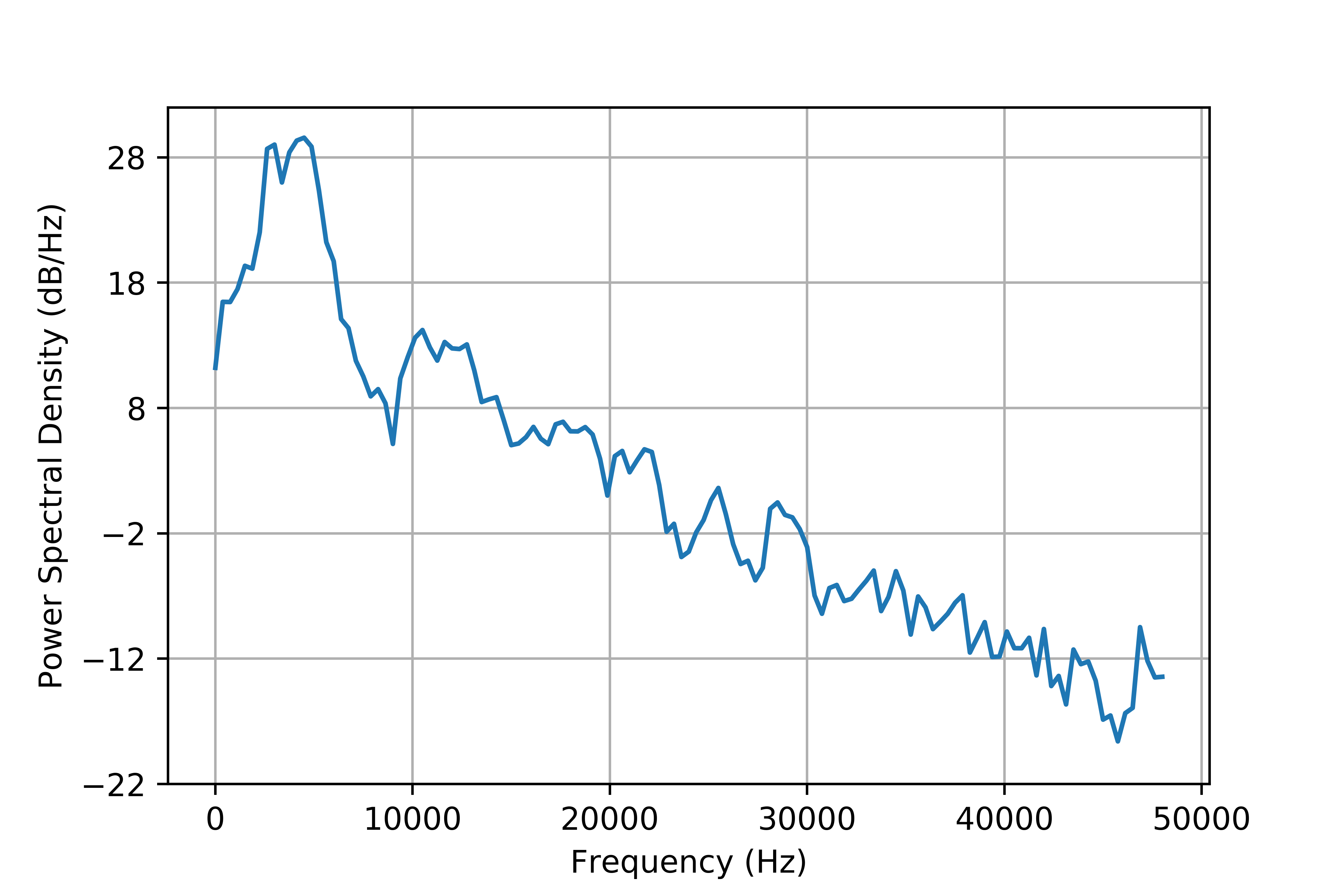}
\caption{Power spectral density versus frequency.}
\label{Figure11}
\end{figure}

\begin{figure}[h!]
\centering
\includegraphics[scale=0.5]{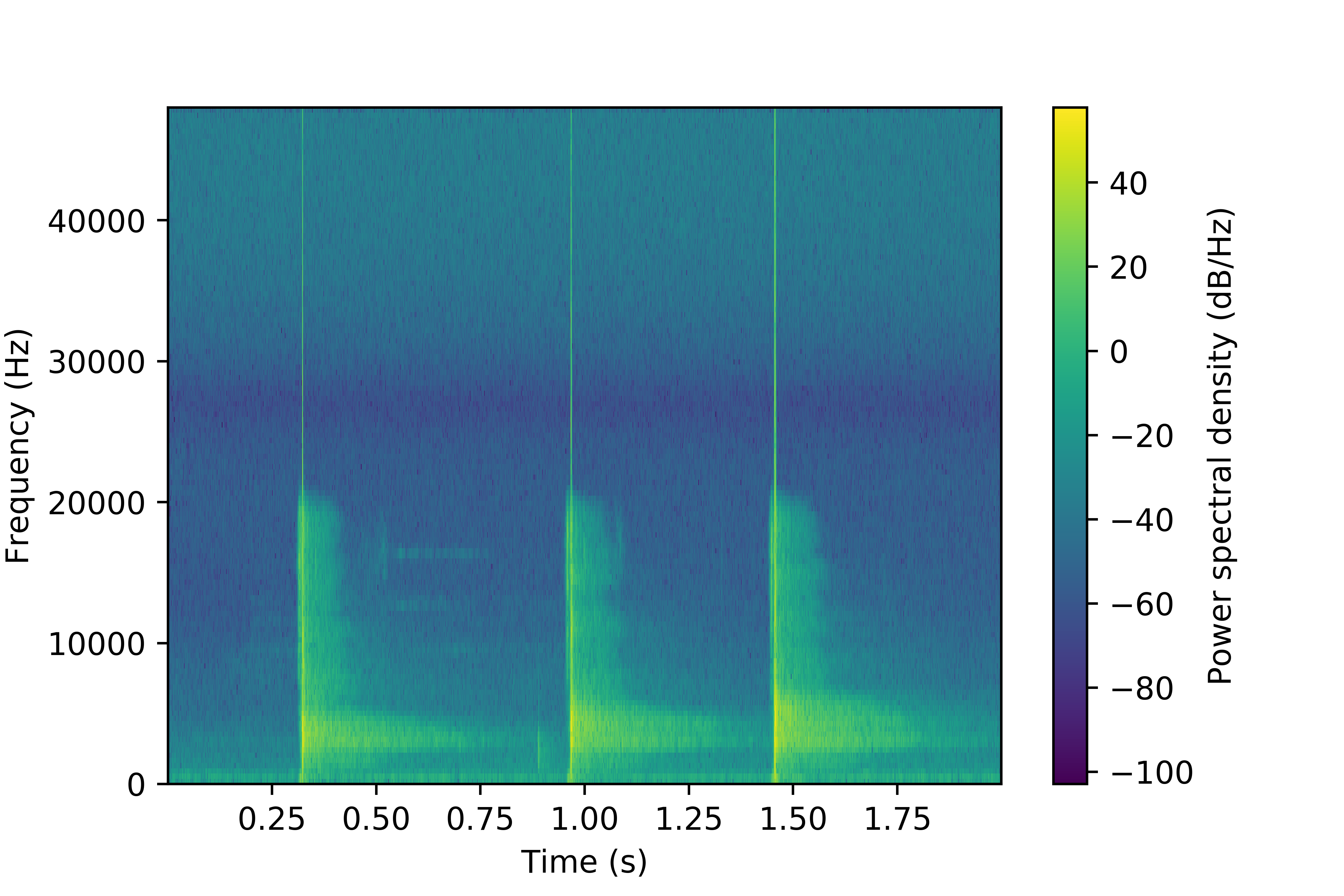}
\caption{Spectrogram of the finger snapping.}
\label{Figure12}
\end{figure}

\section{Impact of FFT Segment Size on Spectrogram Resolution}

The effect of segment division on the results of FFT analysis was investigated using a finger-snapping recording sampled at 44100 Hz. The recording was divided into segments of 1000, 5000, and 50000 points, and the FFTs of each segment were calculated. The resulting PSDs and spectrograms were plotted in Figures 13, 14, and 15, respectively. As the number of segments increased, the resolution of the PSD improved, demonstrating the importance of selecting an appropriate segment size for FFT analysis.

\begin{figure}[h!]
\centering
\includegraphics[scale=0.5]{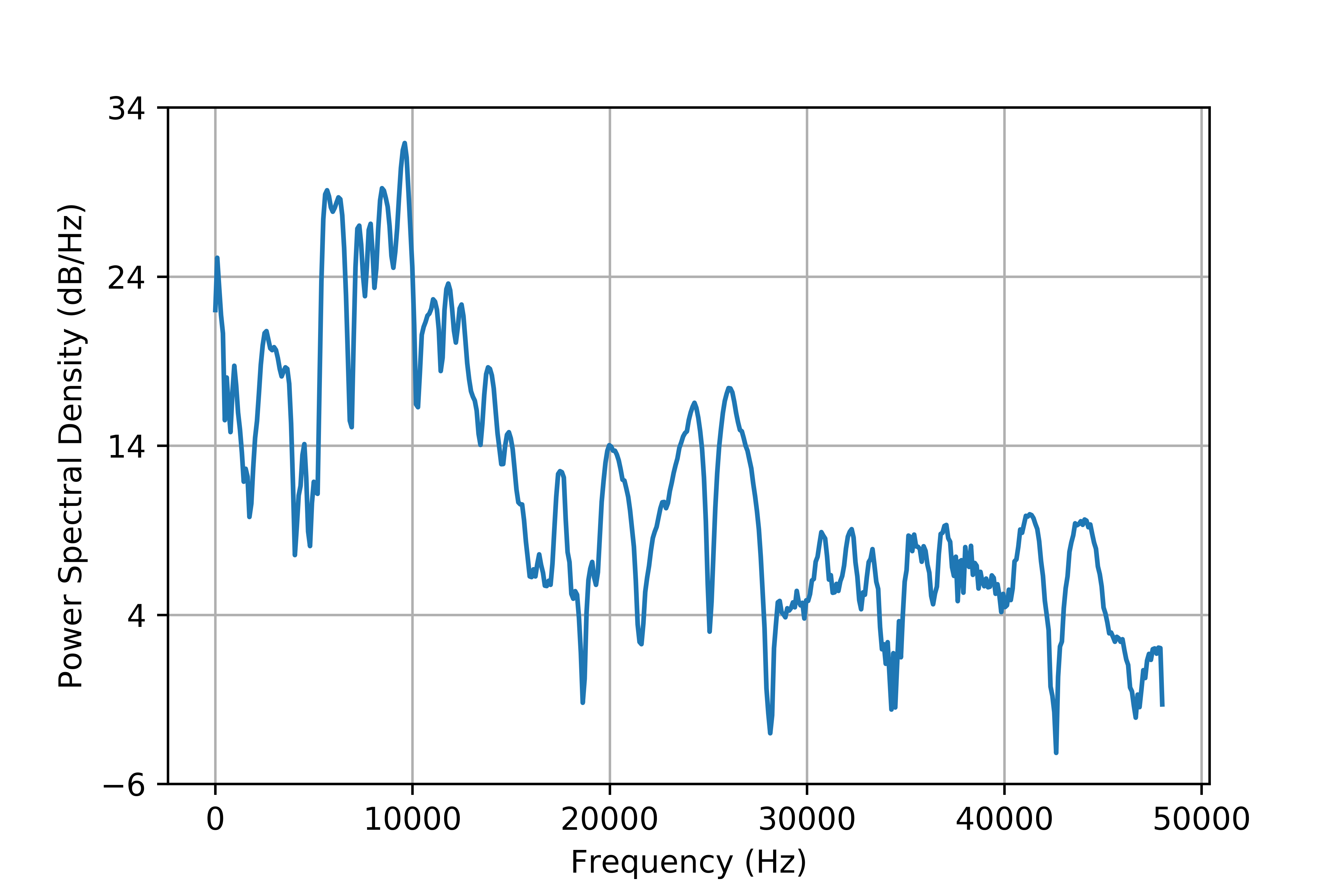}
\caption{Power spectral density versus frequency with 1000 segments.}
\label{Figure13}
\end{figure}

\begin{figure}[h!]
\centering
\includegraphics[scale=0.5]{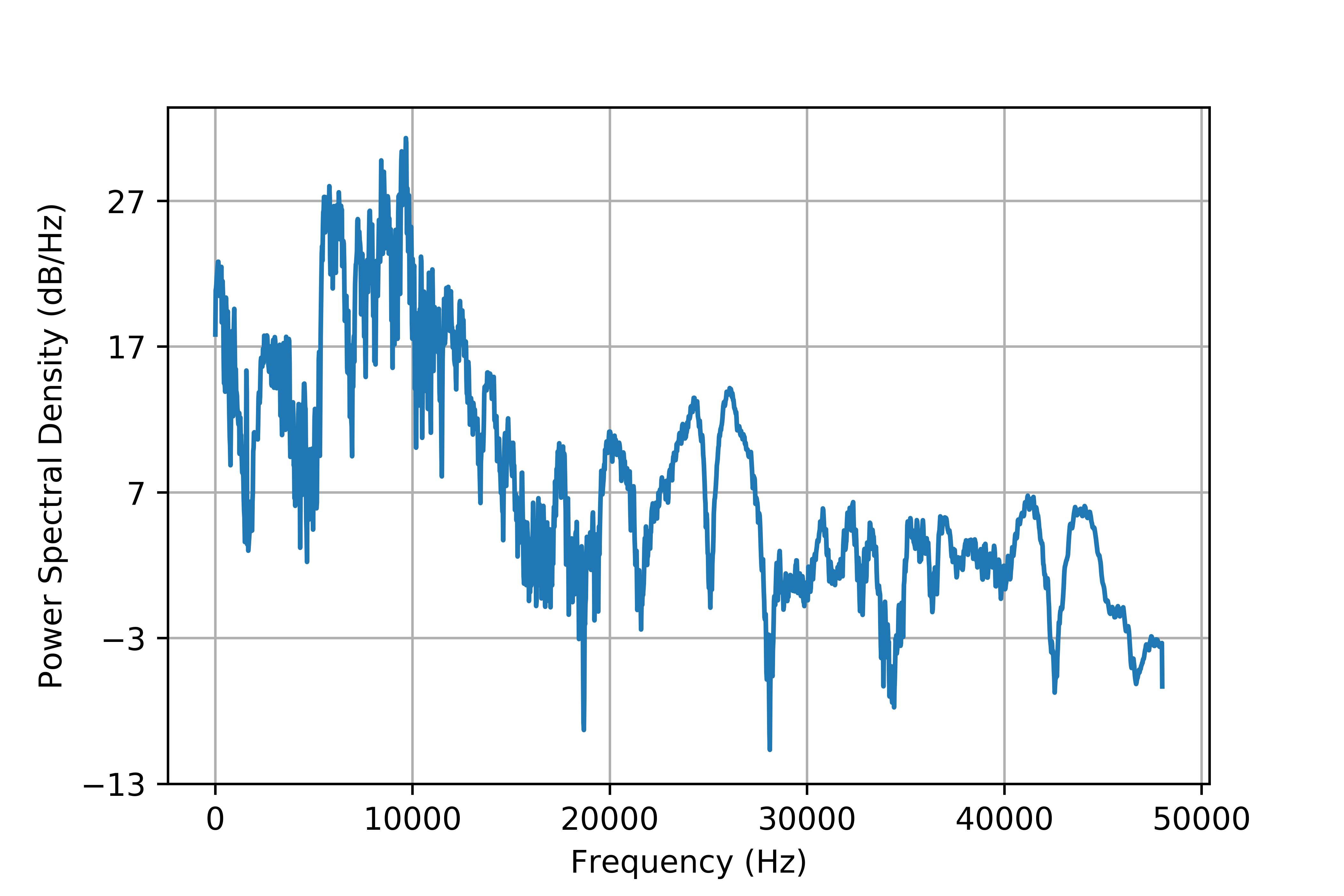}
\caption{Power spectral density versus frequency with 5000 segments.}
\label{Figure14}
\end{figure}

\begin{figure}[h!]
\centering
\includegraphics[scale=0.5]{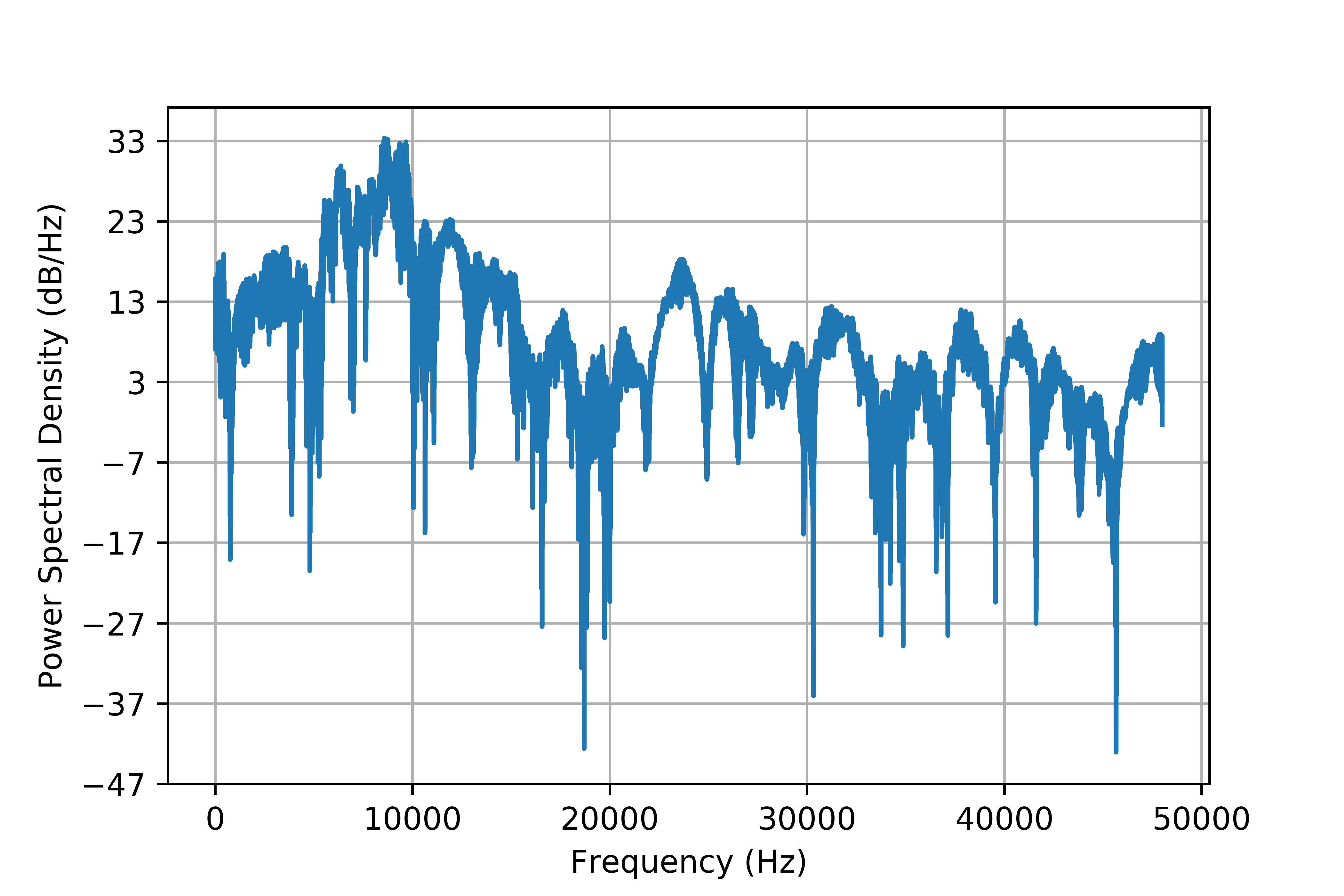}
\caption{Power spectral density versus frequency with 50000 segments.}
\label{Figure15}
\end{figure}

The resolution of spectrograms increases with the number of segments used in the analysis. Figures 16, 17, and 18 show the effect of using 1000, 5000, and 10000 points per segment, respectively. As the segment size increases, the spectrogram's resolution improves, as evident from the distinct rectangles formed in Figures 17 and 18. In Figure 18, the 44100 Hz sampled finger snapping signal, consisting of 192000 data points, is divided into approximately 19 segments of 10000 points each, with 18 of these segments represented in the figure. This demonstrates that the choice of segmentation size for FFT is crucial and should be carefully considered before performing the analysis.

\begin{figure}[h!]
\centering
\includegraphics[scale=0.5]{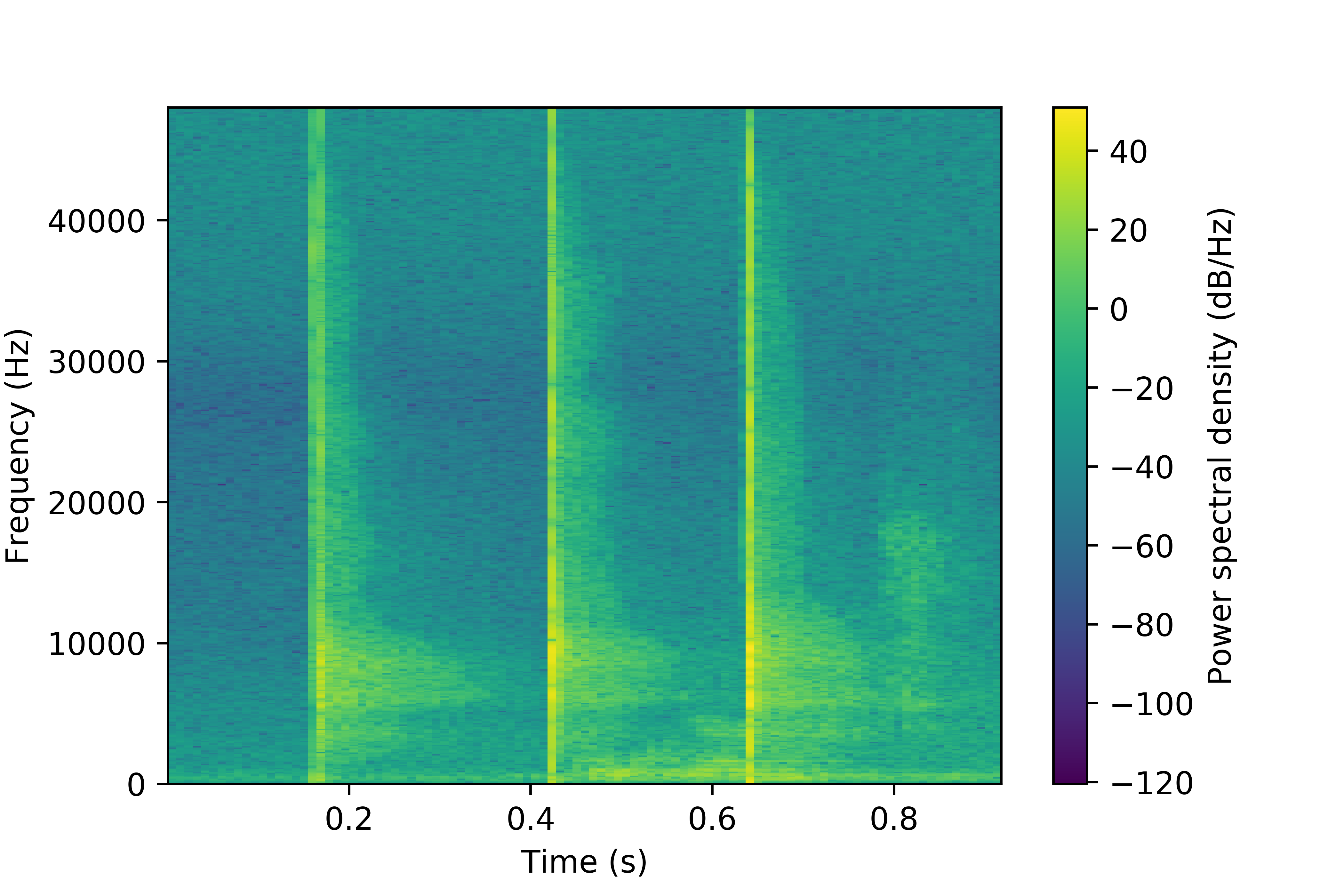}
\caption{Spectrogram of the finger-snapping with 1000 segments.}
\label{Figure16}
\end{figure}

\begin{figure}[h!]
\centering
\includegraphics[scale=0.5]{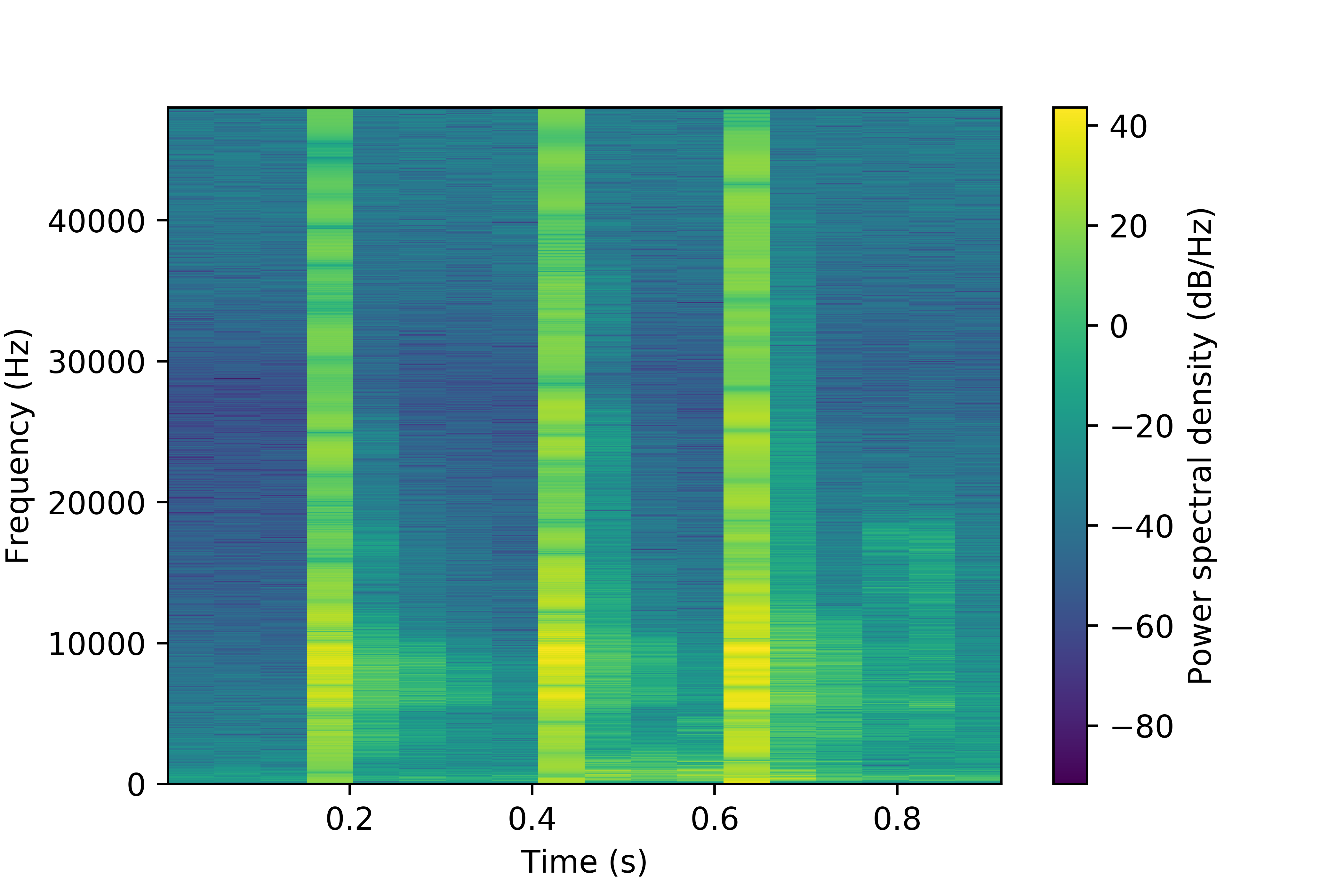}
\caption{Spectrogram of the finger-snapping with 5000 segments.}
\label{Figure17}
\end{figure}

\begin{figure}[H]
\centering
\includegraphics[scale=0.5]{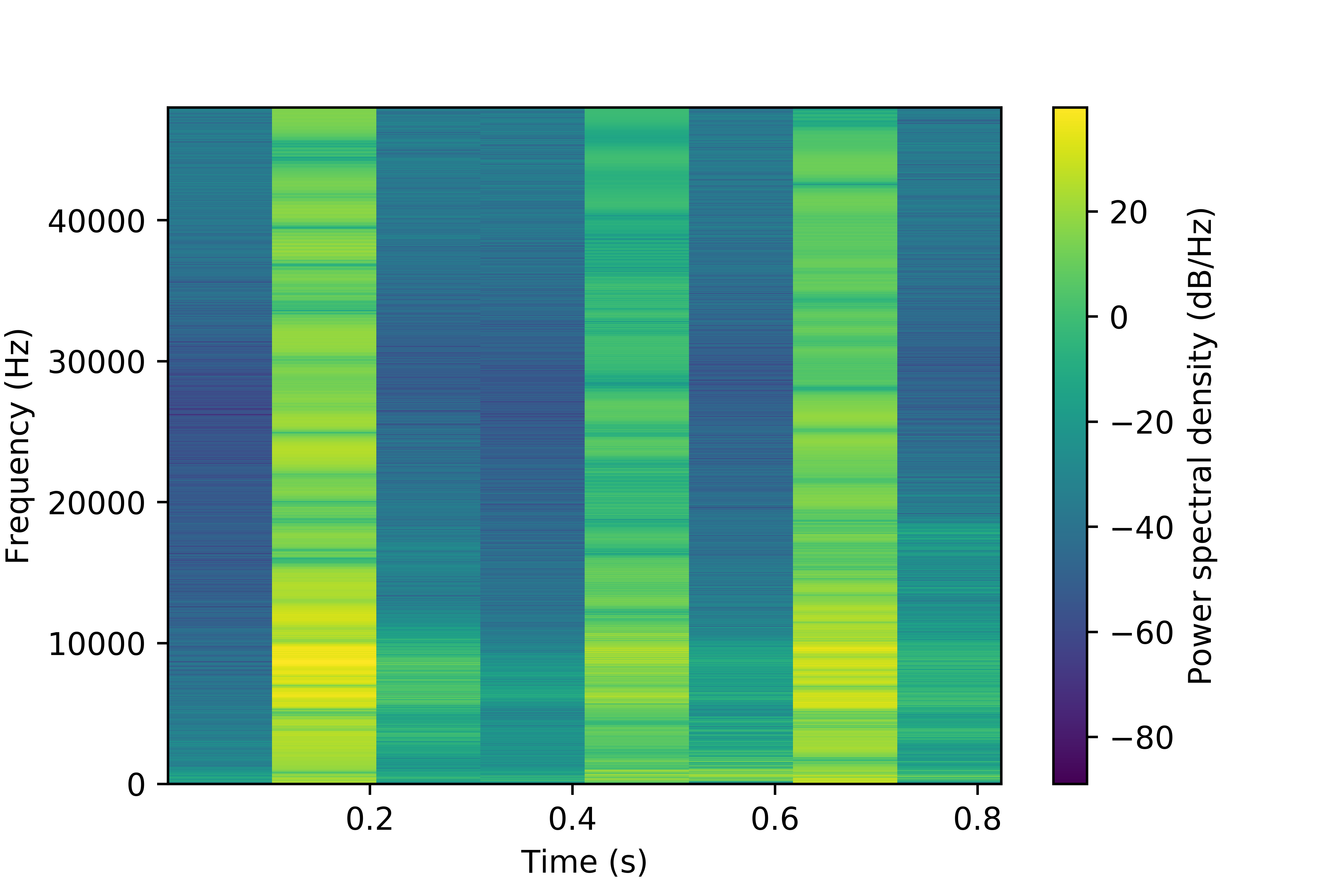}
\caption{Spectrogram of the finger-snapping with 10000 segments.}
\label{Figure18}
\end{figure}






\end{document}